\documentclass{article}
\input epsf.sty
\input colordvi.sty

\newcommand{\bright}{\begin{flushright}}
\newcommand{\eright}{\end{flushright}}
\newcommand{\bminip}{\begin{minipage}}
\newcommand{\eminip}{\end{minipage}}
\newcommand{\bcent}{\begin{center}}
\newcommand{\ecent}{\end{center}}
\newcommand{\beq}{\begin{equation}}
\newcommand{\eeq}{\end{equation}}
\newcommand{\beqa}{\begin{eqnarray}}
\newcommand{\eeqa}{\end{eqnarray}}
\newcommand{\barr}{\begin{array}}
\newcommand{\earr}{\end{array}}

\newcommand{\reflef}{(\ref}
\newcommand{\MP}{M_{\rm P}}

\newcommand{\Lmd}{\Lambda}
\newcommand{\gsim}{\mbox{\raisebox{-.3em}{$\;\stackrel{>}{\sim}\;$}}}
\newcommand{\lsim}{\mbox{\raisebox{-.3em}{$\;\stackrel{<}{\sim}\;$}}}

\begin{document}
\baselineskip=0.6cm

\bcent
{\Large\bf Possible time-variability of the fine-structure constant expected from the accelerating universe
}\\
Yasunori Fujii\\[.0em]
Advanced Research Institute for Science and Engineering, \\[.0em]
Waseda University, 169-8555 Tokyo, Japan\\[.0em]

\ecent
\mbox{}\\[-1.6em]

\baselineskip=0.4cm
\noindent
\bcent
\bminip{14cm}
{\large\bf Abstract}\\[.2em]
We present a theoretical calculation on the time-variability of the fine-structure constant to fit the result of the recent precise analysis of the measurement of the QSO absorption lines.  We find the parameters of the scalar-tensor theory to be determined much more accurately than fitting the accelerating universe itself, but leading not to easy detections of the effect on the equation of state of the dark energy in the earlier epochs. 
{\small 

}
\eminip
\ecent
\mbox{}\\[-1.1em]

\baselineskip=0.6cm

To understand the accelerating universe \cite{accel}, we have developed a cosmological model in the presence of a cosmological constant \cite{cup} based on the scalar-tensor theory (STT) in its simplest version.  The theoretical analysis consists of the two steps.

In Step I, we derive solutions on the spatially flat Friedmann-Robertson-Walker  universe also with the assumed spatially uniform scalar field $\phi$, featuring no inflationary expansion in spite of a ``large" dimensionful constant $\Lmd$ which we simply added to STT in its Jordan conformal frame (JCF).  As one of the most remarkable achievements at this stage, we have successfully implemented the ``scenario of the decaying cosmological constant," allowing us to understand the well-known number $10^{-120}$ simply as $t_{0}^{-2}$, where $t_0$ is the present age of the universe, $\sim 10^{60}$ in the reduced Planckian unit system with $c=\hbar =\MP(=(8\pi G_0)^{-1/2})=1$ with $G_0$ for today's value of Newton's constant.  This conclusion is reached in the Einstein conformal frame (ECF) which is identified as the physical conformal frame.  The scalar field which is canonical here is denoted by $\sigma$, behaving as dark energy.  As we also emphasize, the exponential potential of $\sigma$, a favored choice in the phenomenological approach of quintessence  \cite{quint}, is an automatic outcome from our theoretically constrained model.

There are two issues to be discussed.  First we argue that the masses of matter fields should be time-independent in the physical conformal frame, requiring the traditional Brans-Dicke ``model" in JCF be replaced by another model, now featuring global scale-invariance except for the $\Lmd$ term.  Secondly, the density of $\sigma$ falls off in the same way as the matter density.  This ``scaling behavior" must be replaced by the ``tracking behavior" in order for the universe to be in fact accelerated.  This situation is realized if the $\sigma$-energy stays nearly constant for some duration of time.

We implemented this in Step II by introducing another scalar field $\chi$, which provides a potential to which $\sigma$ is temporarily trapped.  The field $\sigma$ goes down oscillating toward one of the minima of the potential.  This oscillation is so small that it hardly shows itself in the process of acceleration, but it does in the temporal variation of the fine-structure ``constant," through the relation in ECF \cite{cup};
\beq
\frac{\Delta\alpha}{\alpha}={\cal Z}\frac{\alpha}{4\pi}\zeta\Delta\sigma \approx 4.6\times 10^{-3} \Delta\sigma,
\label{tvalpha_1}
\eeq
with $\Delta$ always understood for a difference from today's value.  Also ${\cal Z}=5$ is for the effective number of unit charge of the fundamental quarks and leptons in the loop, while $\zeta(=6+4\omega)=1.5823$ was used for our fitting the acceleration.  To avoid a conflict with the result of solar-system experiments, $\zeta \lsim 10^{-3} $\cite{bert}, we assume a finite force-range of $\sigma$.  We halved the numerical coefficient in (6.194) of \cite{cup}, because we had mistakenly included contributions from the both sides of the photon self-energy diagram.  Since $\Delta\alpha/\alpha$ is proportional to the change of $\sigma$, the former also oscillates, inheriting the crucial physical process  in the cosmological acceleration.

The above equation \reflef{tvalpha_1}) is unique in the sense that we derived it based on the standard minimal electromagnetic interaction of charged fields.  We did not make any {\em ad hoc} assumption like the coupling in the form of $F(\phi)F_{\mu\nu}F^{\mu\nu}$ \cite{bek}, based on which many phenomenological analyses have been attempted on the time-variability of $\alpha$ \cite{mnap,nunav}.  We point out that we used the quantum-anomaly technique in the relativistic quantum field theory to implement the idea of spontaneous breaking of scale invariance in JCF as mentioned before.

Recently Levshakov et al reported a nonzero time-dependence of the fine-structure constant $\alpha$ after a careful analysis of the observed QSO absorption lines, based on the method of Single Ion Differential $\alpha$ Measurement; $(\Delta\alpha/\alpha)\times 10^6= -0.12\pm 1.79$ and $5.66\pm 2.67$ at $z=1.15$ and $1.84$, respectively \cite{Levs,prv,bdh}, corrected also for the improved estimate of the Fe\hspace{.2em}{\footnotesize II} sensitivity coefficients \cite{porsev}.  We note that the quoted errors include both statistical and systematic uncertainties.   Taking a rather tentative nature of their analysis into account, we still believe it worth attempting to apply a theoretical estimate \reflef{tvalpha_1}) to their result, going beyond our earlier phenomenological analyses \cite{yfsm} on the then available observational results \cite{qso}.  We also include the constraint obtained from the Oklo natural reactors \cite{oklo} as well as the laboratory measurement of $\dot{\alpha}/\alpha$ \cite{peik}.  In this way we expect to constrain the parameters and initial values much better than fitting the accelerating universe itself.  It seems unlikely, on the other hand, that this type of solution results in the past-time deviation of the equation of state $w=p/\rho$ for the dark energy from $-1$ at any practical level of detection, indicating probably an important insight into what behavior of $w$ is going to look like.

As we admit, however, we have never swept the entire space of 5 parameters and 4 initial values, having shown the fit in Fig. 5.8 of \cite{cup} only to illustrate an explicit solution that reasonably fits the data for the accelerating universe.  This solution, to be referred to as the ``reference" solution, turns out, however, to predict too large values on $\Delta\alpha/\alpha$ to be compared with the recent observation \cite{Levs}-\cite{bdh}.  As a remedy we change the value of the parameter $\gamma$, a coefficient in front of the sine-Gordon term in the potential in (5.58) of \cite{cup}, from 0.8 to 0.81.  Corresponding changes of other parameters and initial values are to be in order.  In view of the nonlinear nature of our cosmological equations, however, we follow a cautious approach by starting with the reference solution, with the parameters ($\Lmd =1, \zeta=1.5823, m=4.75, \gamma =0.8, \kappa =10.0$) and the initial values ($\sigma_1=6.7544, \sigma_1'=0, \chi_1=0.21, \chi_1'=0$) imposed at $t_1=10^{10}$ (sometime after the primordial inflation), also in the reduced Planckian unit system.  We search only the neighborhood of the reference solution except for $\gamma$.

The solutions we reached in this heuristic manner are shown in Fig. 1, with two fixed parameters $\rho$ and $\delta x$ and the initial value $\sigma_1$ varied in the range including the reference value 6.7544.  Let us begin with short explanations  on the proposed constants $\rho$ and $\delta x$.
\mbox{}\\[-1.9em]

We are going to identify the epoch about $1.9\times 10^9$ years ago for the Oklo phenomenon with a zero of the oscillating $\Delta\alpha/\alpha$.  This is affected obviously by a factor $\rho$ chosen to give the age of the universe $t_0 = \rho \times 1.37 \times 10^{10}{\rm y}$.  We also notice that the calculated curves for $\Delta\alpha/\alpha$ will be shifted horizontally in Fig. 1, if we change the initial time $t_1$.  More technically, our cosmological equations were integrated with respect to $x=\log_{10} t$ starting at the initial value $x_1$ chosen conveniently to be 10.  By changing $x_1$ by a small amount $\delta x$, we are allowed to {\em re-use} the result of the reference solution but with $\gamma =0.81$, resulting in the approximate shifting the curves in proportion to $\delta x$.  From an overall fit to the Oklo at around $z\approx 0.16$ together with two QSO data, we favor the choice $\rho=1.02$ and $\delta x=0.05$ as reasonable fine-tunings.  The former choice for $t_0 \approx 1.40\times 10^{10}{\rm y}$ will be tolerated.

\mbox{}\\[-14.7em]
\hspace*{0em}
\bminip{16cm}
\epsfxsize=14cm
\epsffile{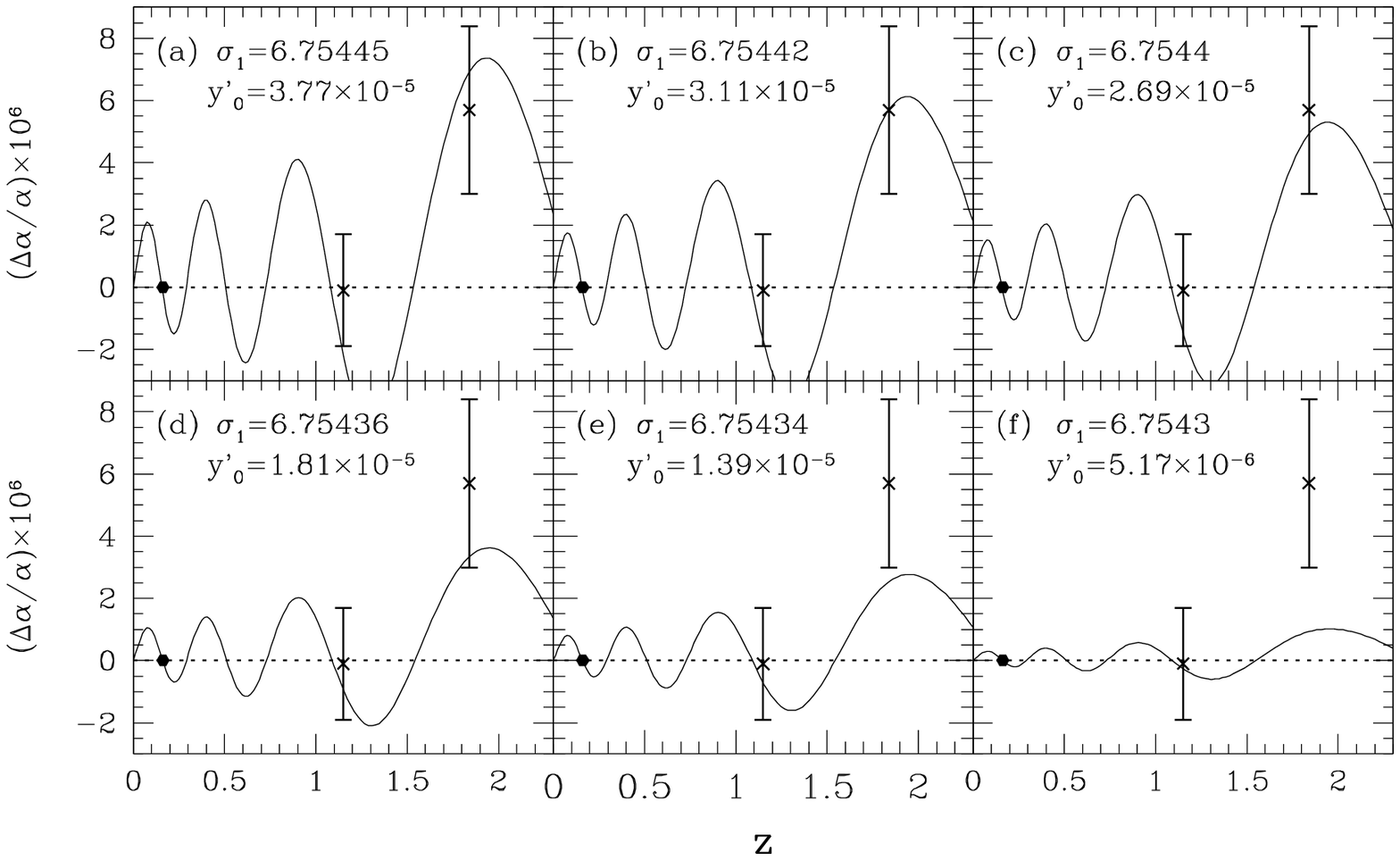}
\eminip
\mbox{}\\[-2.9em]
\begin{figure}[h]
\caption{$(\Delta\alpha/\alpha)\times 10^6 $ is plotted against the redshift $z$ for different initial values $\sigma_1$ as shown in each entry, to be compared with the QSO data \cite{Levs}-\cite{bdh}. Also $y'_0 \equiv (d(\Delta\alpha/\alpha)/dz)_{z=0}$, the slope of the calculated curve at $z=0$, is to be compared  with the laboratory measurement of $\dot{\alpha}/\alpha$ \cite{peik}.  The Oklo constraint \cite{oklo} is marked by a blob at $z\approx 0.16$.
}

\end{figure}

Note also that the Oklo phenomenon gives an upper-bound of $|\Delta\alpha/\alpha| \lsim 10^{-7}$, or even smaller $\lsim 10^{-8}$ \cite{oklo}, represented by a blob around $z\approx 0.16$, with a nearly invisible error bar in this plot.

Given estimates  of $\rho$ and $\delta x$ as above, we now vary $\sigma_1$, which might be considered as a representative of other initial values or parameters.  In Fig. 1, we find quite a variety of the behavior of $y(z)=\Delta\alpha/\alpha$ for different values of $\sigma_1$ in a rather short interval of 0.00015 including the reference value 6.7544.

We also note that the slope of the curve $y'_0$ at the origin is related to the result of the laboratory measurement of $(\dot{\alpha}/\alpha)_0$ at the present epoch by
\beq
\left( \frac{\dot{\alpha}}{\alpha} \right)_0 =-H_0 y'_0,
\label{alphadot_1}
\eeq
as given most recently by $(-0.26 \pm 0.39)\times 10^{-15}{\rm y}^{-1}$ \cite{peik}, which can be translated into $y'_0 =(3.5 \pm 5.2)\times 10^{-6}$ for $h=0.73$, also  to be denoted by $y'_{\rm lbt}\pm \sigma_{\rm lbt}$.\\[-1.6em]

As we find, the computed curves agree reasonably well with the two QSO data at $z=1.15$ and $z=1.84$ for $\sigma_1 \gsim 6.7544$, which happens to be the reference value in the solution (c),  but with unacceptably large $y'_0$  with $v\equiv (y'_0 -y'_{\rm lbt})/\sigma_{\rm lbt} \gsim 4.5$.  The deviation $v$ is larger for larger $\sigma_1$, while it is smaller for smaller $\sigma_1$, coming down to $ v=0.3$ in (f), for example, but now with an obviously poor agreement with the QSO data.  In spite of a desperately necessary effort to search for wider class of solutions, also given a still small number of precise measurements particularly on the QSO result, the very presence of ``compromises," exemplified by the solutions (d) or (e), with $v=2.8$ and 2.0, respectively, seems encouraging since, as we show shortly, all the solutions discussed here share nearly the same cosmological consequences.

More explicitly, we notice that Figs. 2 and 3,  corresponding to Figs. 5.8 and 5.9 in \cite{cup}, respectively, on the cosmological behaviors computed from the typical fit (e) in Fig. 1,  remain nearly the same for all other solutions (a)-(f); no discernible differences are recognized for their results.  This is a general tendency, due to the fact that considerable difference in $\Delta\alpha/\alpha$ comes from the same in $\sigma$, which is only a ``small" ripple if viewed from the overall cosmological behaviors, as shown in these plots.

\mbox{}\\[-2.5em]
\bminip{7.5cm}
\mbox{}\\[-.2em]
\epsfxsize=6.5cm
\epsffile{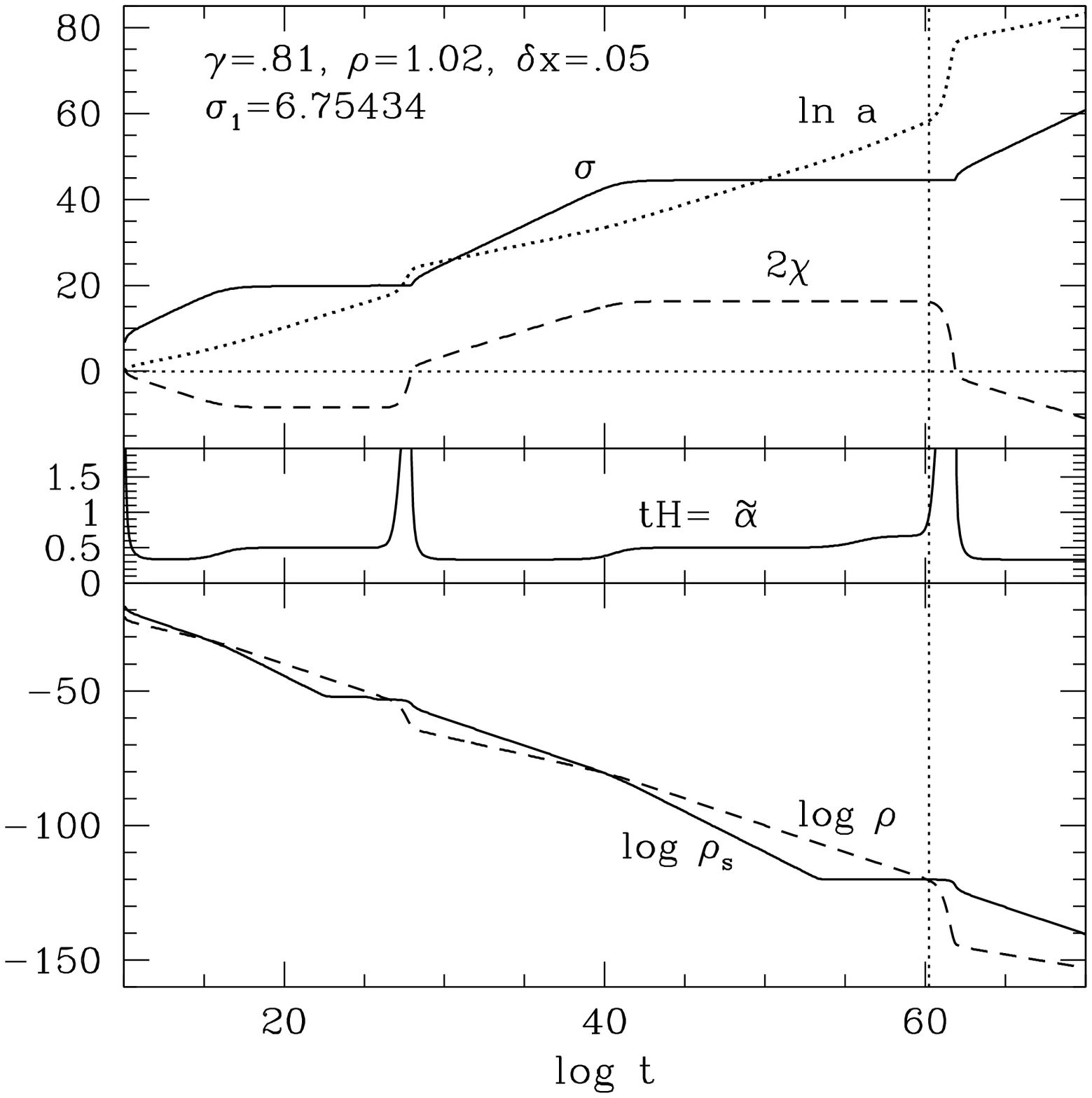}
\mbox{}\\[-1.0em]

\noindent
Figure 2: Cosmological behaviors in ECF for the solution (e) of Figure 1, with the reference counterpart in Fig. 5.8 of \cite{cup},  $\rho$ and $\rho_s$ for the densities of the matter and the $\sigma\chi$ system (dark energy), respectively. Present epoch, $\log_{10}t_0=60.20$, is shown by a vertical dotted line.
\eminip
\hspace{1.8em}
\bminip{7.5cm}
\mbox{}\\[-.6em]
\epsfxsize=6.5cm
\epsffile{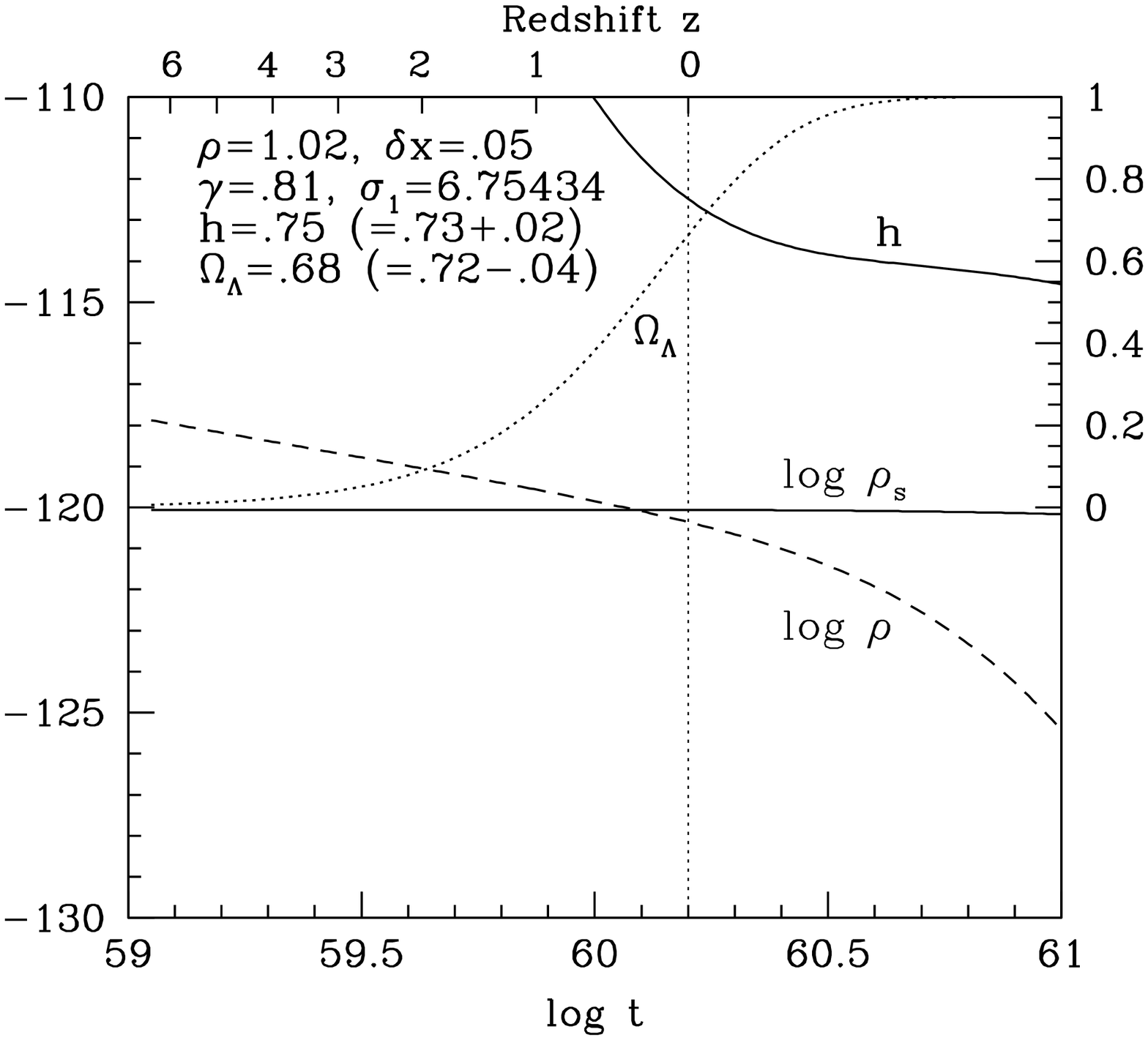}
\mbox{}\\[-1.5em]

\noindent
Figure 3: Magnified view of Figure 2 near the present epoch, corresponding to Fig. 5.9 of \cite{cup}. Today's values of $h$ and $\Omega_\Lmd$ are also shown, measured according to the right-hand vertical scale, found within the observed ranges \cite{wmap}.
\eminip
\mbox{}\\[-.0em]

In fact, corresponding to Fig. 5.10 of \cite{cup}, we have Fig. 4 with the vertical scale magnified by as much as 10,000 times compared with that of the upper panel of Fig. 2.  Measuring $\Delta\alpha/\alpha$ is something like watching a portion of $\sigma$ around the present epoch, featuring an apparently smooth and flat section followed by a sudden but tiny step-like rise, in the upper panel of Fig. 2, through a microscope with much better resolution.  In other words, $\Delta\alpha/\alpha$, if confirmed, is expected to probe the parameters much more accurately than the cosmological acceleration can do.  Just for the sake of comparison, we show the corresponding reference solution in Fig. 5, basically the same as Fig. 5.10 of \cite{cup}.

\mbox{}\\[-1.2em]
\bminip{7.5cm}
\mbox{}\\[-.2em]
\epsfxsize=6.5cm
\epsffile{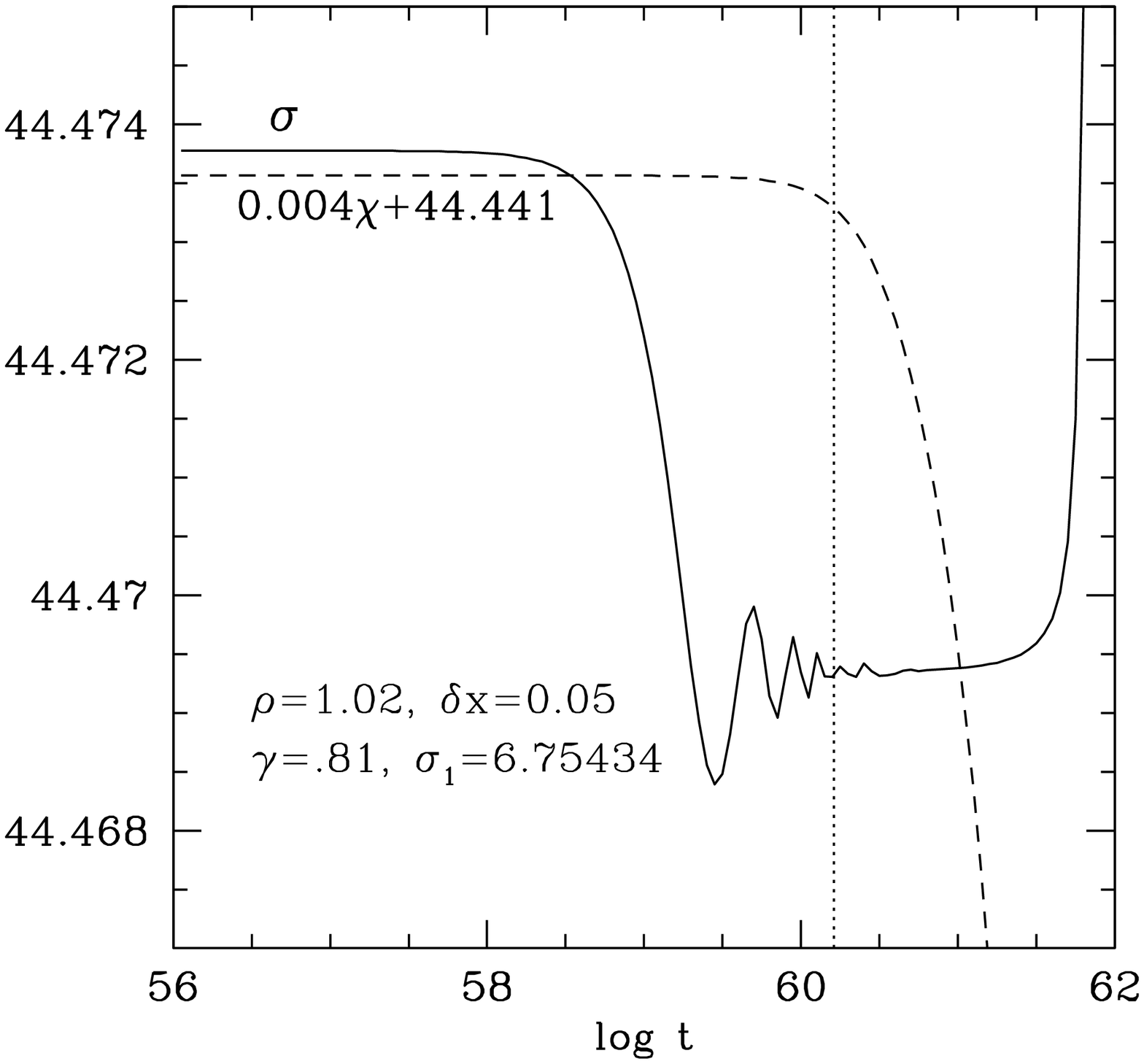}
\mbox{}\\[-1.3em]

\noindent
Figure 4: Magnified view of the upper panel of Fig. 2, with the vertical scale magnified by about 10,000 times.  $\sigma$ is being trapped by the potential provided by $\chi$, which starts falling toward zero, thus releasing $\sigma$ eventually.
\eminip
\hspace{2.1em}
\bminip{7.5cm}
\mbox{}\\[-1.1em]
\epsfxsize=6.5cm
\epsffile{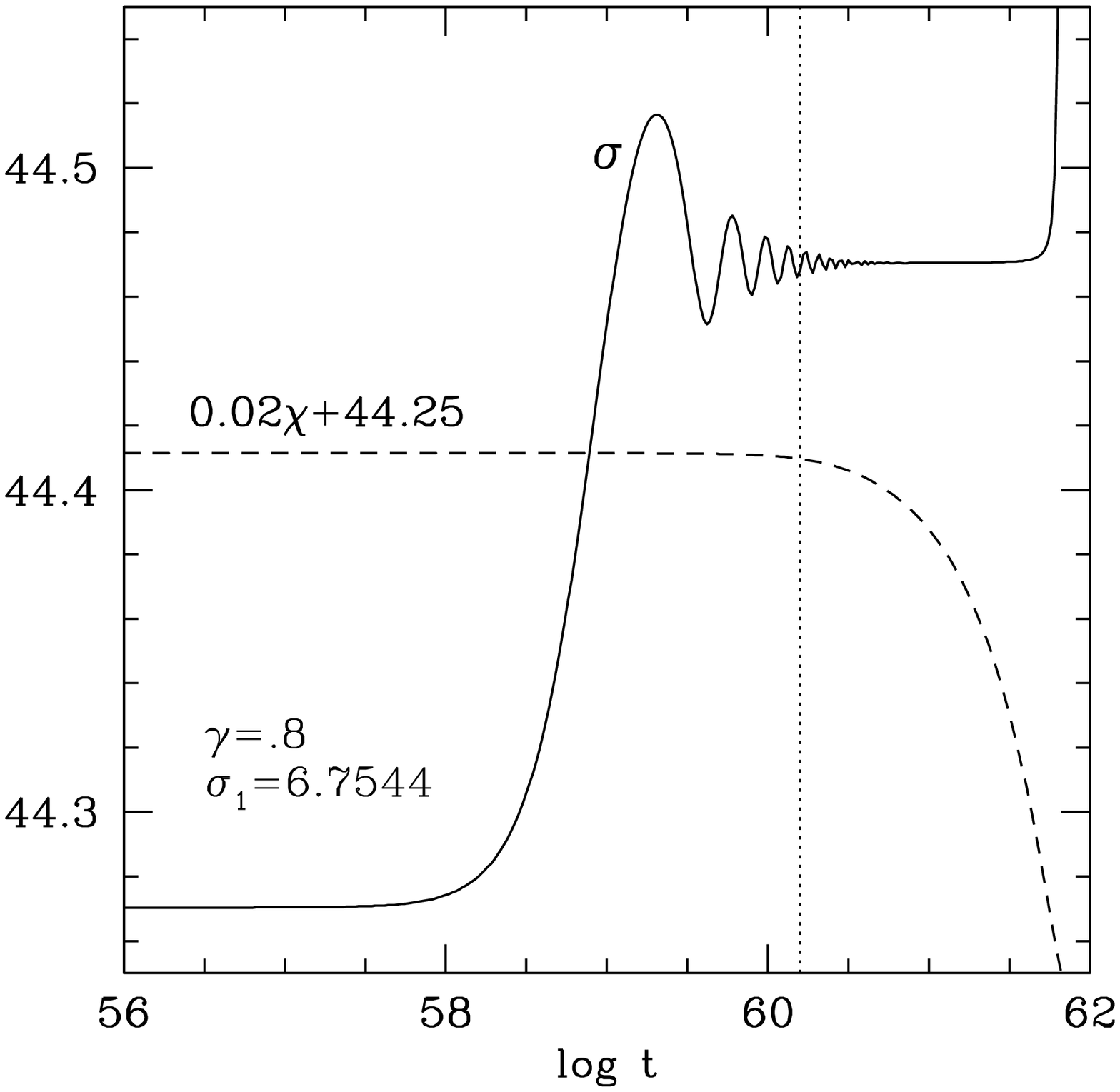}
\mbox{}\\[-1.6em]

\noindent
Figure 5: The reference solution counterpart of Fig. 4.  The behaviors may look similar, but the rate of vertical magnification is roughly 30 times less than in Fig. 4.  
\eminip
\mbox{}\\[.8em]

The difference between Figs. 4 and 5 can be understood by different values of $\sigma$ when it has ``cruised" before it is captured by the potential due to $\chi$.  We simply say that $\sigma$ in Fig. 4 ``happened" to be located much closer  to a potential minimum than in Fig. 5.  This is the reason why the  ripple of $\sigma$ is sufficiently small to be in agreement with the small observed $\Delta\alpha/\alpha$.   In other words the reference solution would have predicted $\Delta\alpha/\alpha$ an order larger than the current observation \cite{Levs}-\cite{bdh}.

Figures 4 and 5 also suggest that simple behaviors of $\sigma$ in the immediate past might have left some of observable effects.  In principle, the oscillation of $\sigma$ entails a more or less oscillatory behavior of $w=p_s/\rho_s =(K-V)/(K+V)$, with $K=\dot{\sigma}^2/2 +\dot{\chi}^2/2$, while $V$ is given by (5.58) of \cite{cup} for the interaction energy of the system of  $\sigma$ and $\chi$ which together comprise dark energy, denoted by the subscript $s$, in Step II in our theoretical model.  Unfortunately, the relationship between the two quantities appears to be complicated  as suggested by comparing Figs. 1 and 6, partly because $\chi$ is also involved.  Furthermore the magnitude of time-variability of $w$ is much smaller than in smooth behaviors which are expected on phenomenological bases \cite{nunav,lind25,dage}.  This is a general feature shared by other solutions in Fig. 1 as well.  We learn that apparently structureless behavior of $w$ close to $-1$ in a range of small $z$ may not necessarily imply a purely constant $\Lmd$.  It can still be consistent with a nonzero $\Delta\alpha/\alpha$ of the order of $10^{-6}$.

This apparently flat behavior stuck closely to $-1$  in a limited resolution is shown to continue down to $t\sim 10^{54}$ at which $\rho_s$ begins to rise in the past direction, as we illustrate in Fig. 7.  This epoch nearly coincides with the ``equal time," or the time toward the end of CMB epoch.  Nearly at the same time, the effective power exponent $\tilde{\alpha} \equiv tH$, allowing a locally approximate dependence $a\sim t^{\tilde{\alpha}}$, begins, in the future direction, to change from $1/2$ to $2/3$.  In fact we adjusted the parameters and initial values in such a way that the condition $\rho_s \ll \rho$ is met around $t\sim 10^{45}$, thus maintaining the success of the primordial nucleosynthesis undisturbed by a cosmological constant, as also seen in the lower panel of Fig. 2.  It may not be easy to change this feature expected from an overall point of view.  We still find it likely that the smallness of the ripple of $\sigma$ is responsible for near undetectability of the $w$ variability, as we inspect the reference counterpart in Fig. 8.

\hspace{8em}
\bminip{7.5cm}
\mbox{}\\[-1.1em]
\epsfxsize=6.5cm
\epsffile{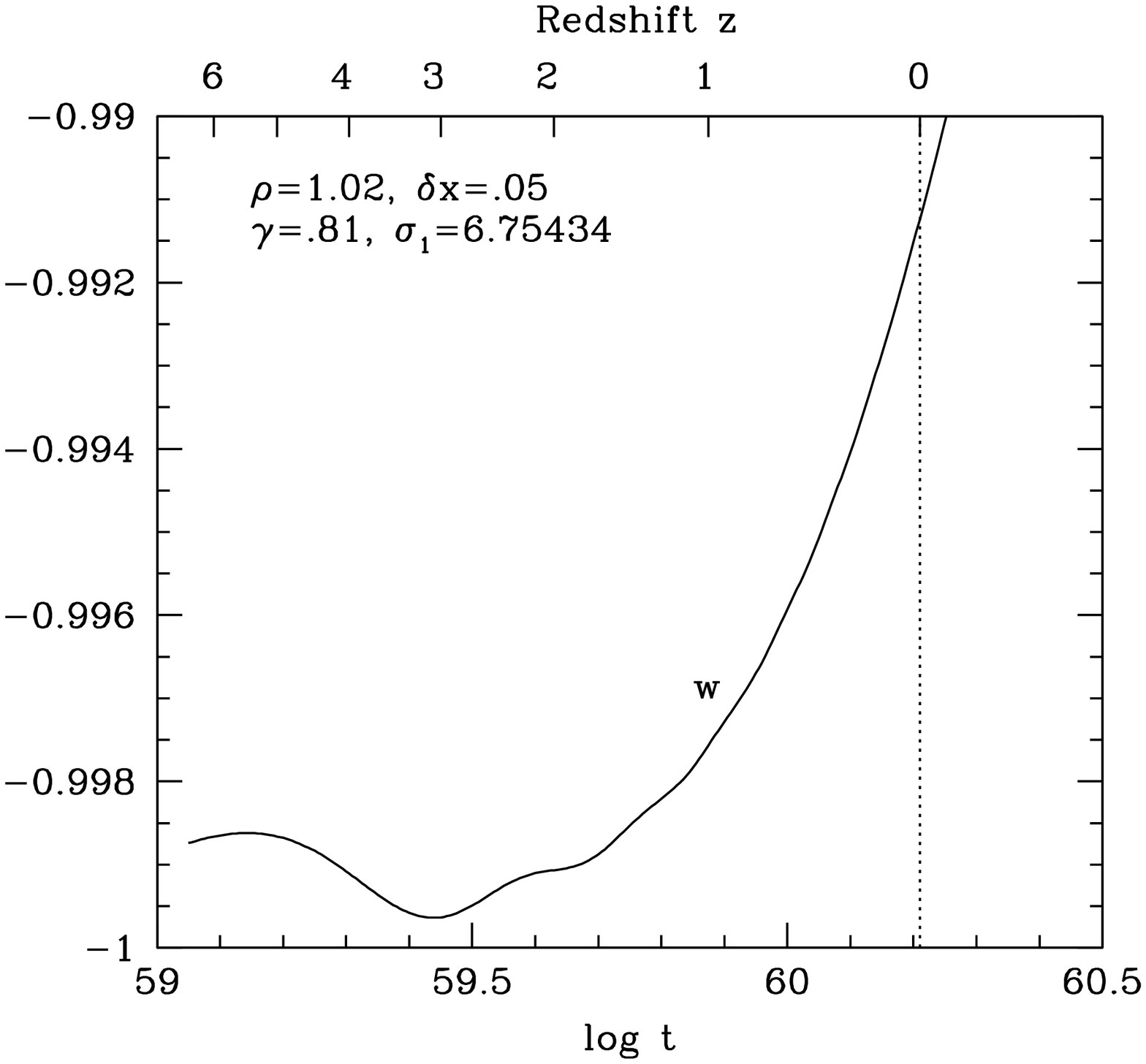}
\mbox{}\\[-.6em]
\eminip
\mbox{}\\[-2.4em]
\setcounter{figure}{5}
\begin{figure}[h]
\caption{The equation of state $w=p_s/\rho_s$ for the solution (e) of Fig. 1 stays so close to the lower bound $-1$ that no detection appears feasible for reasonable range of $z$.  Basically the same result follows for any of other solutions displayed in Fig. 1.
}
\end{figure}
\mbox{}\\[-1.1em]

Unlike in Fig. 7, we do find a number of spike-like behaviors of $w$.  The biggest one centered at $\log_{10}t\sim 58.95$, also including smaller ones toward $\log_{10}t\sim 60$, is due to the slight change of $\rho_s$, as also shown in Fig. 8.  The same type of behavior of $\rho_s$ does not seem an uncommon occurrence among many examples of solution, though rarely occurring for such low values of $z\lsim 2$, or $\log_{10}t \gsim 59.6$.  It might be rather accidental to find no such spikes, though allowing to be much closer to $-1$, in the solutions shown in Fig. 1.  The behaviors of the sharp rise (in the past direction) to $w=+1$ around $\log_{10}t \sim 54$ are nearly the same as in Fig. 7.

Perhaps more important is to note that major behaviors, like the above examples, but unlike those discussed in \cite{nunav,lind25} from phenomenological points of view, may show the features having something in common with a delta function or a step function, to which a power series expansion is known to be ineffective.  This might be a unique conclusion we draw from our theoretical model designed primarily to understand the accelerating universe.  We should be prepared, however,  at least for the possible occurrence of such ``singular" terms.

\noindent
\bminip{7.5cm}
\mbox{}\\[-.2em]
\epsfxsize=6.5cm
\epsffile{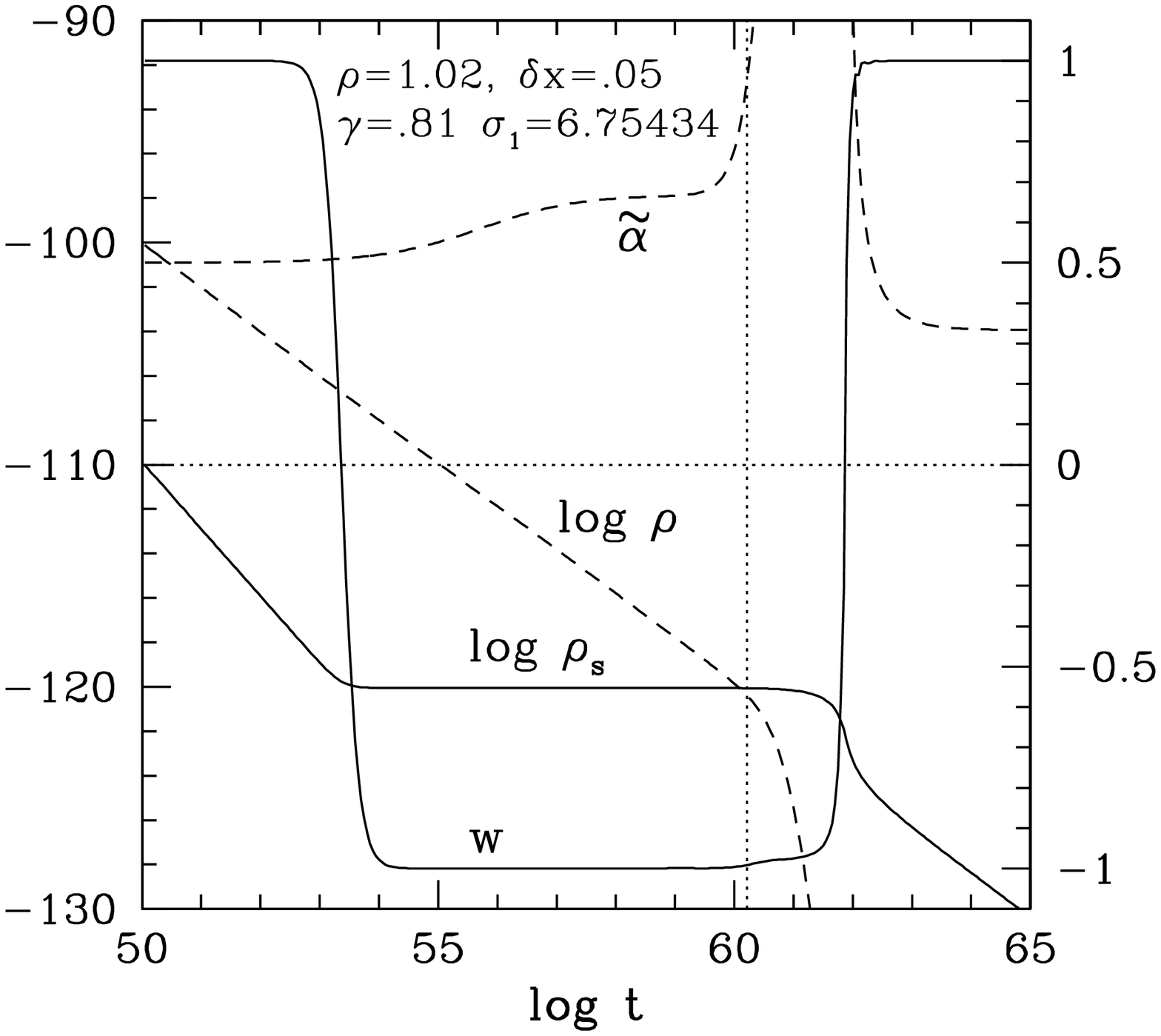}
\mbox{}\\[-.0em]

\noindent
Figure 7: Inter-relations among the matter density $\rho$, the $\sigma\chi$ system density $\rho_s$ (dark energy), measured according to the left-hand scale, while the equation of state for the scalar fields $w$, and the effective power exponent $\tilde{\alpha}$ of the scale factor, allowing $a\sim t^{\tilde{\alpha}}$, measured against the right-hand scale, for the solution (e) in Fig. 1.
\eminip
\hspace{2.1em}
\bminip{7.5cm}
\mbox{}\\[-5.4em]
\epsfxsize=6.5cm
\epsffile{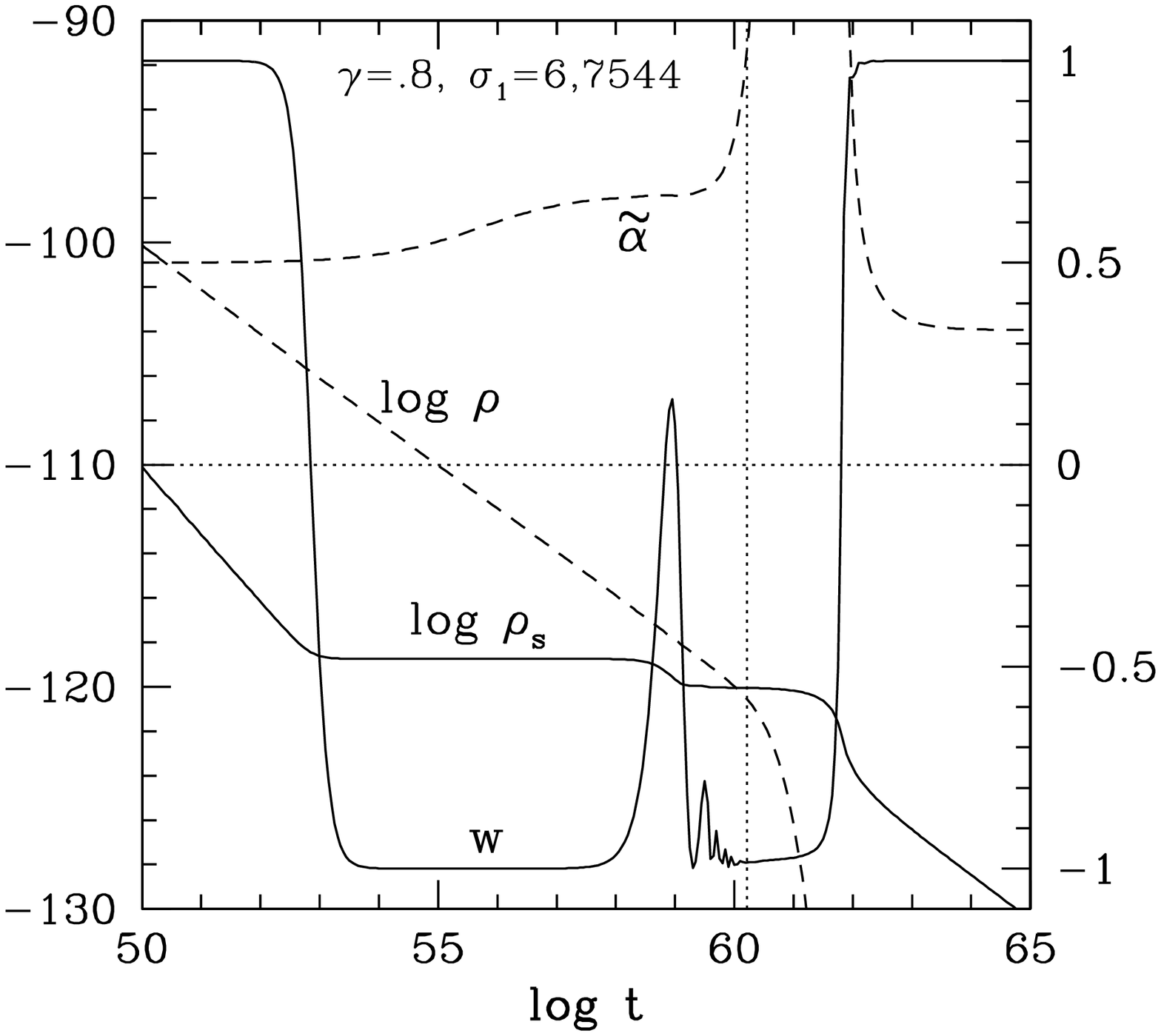}
\mbox{}\\[-.0em]

\noindent
Figure 8: The same as in Fig. 7, but for the reference solution.  Note the occurrence of considerable time-dependence of $w$, unlike in Fig. 7.

\eminip
\mbox{}\\[.6em]

As we emphasized before, we are not fully satisfied by the solutions obtained in  our limited and heuristic approach.  Also the future re-analysis of the QSO measurements may reveal aspects different from those in \cite{Levs}.  Observational searches on the equation of state available at present, on the other hand, appear to be consistent with nearly anything including purely $w=-1$.  Taking these circumstances into account, we should make further efforts to search for other types of solution beyond the neighborhood of the reference solution.

\hspace{8em}
\bminip{7.5cm}
\mbox{}\\[-2.1em]
\epsfxsize=6.5cm
\epsffile{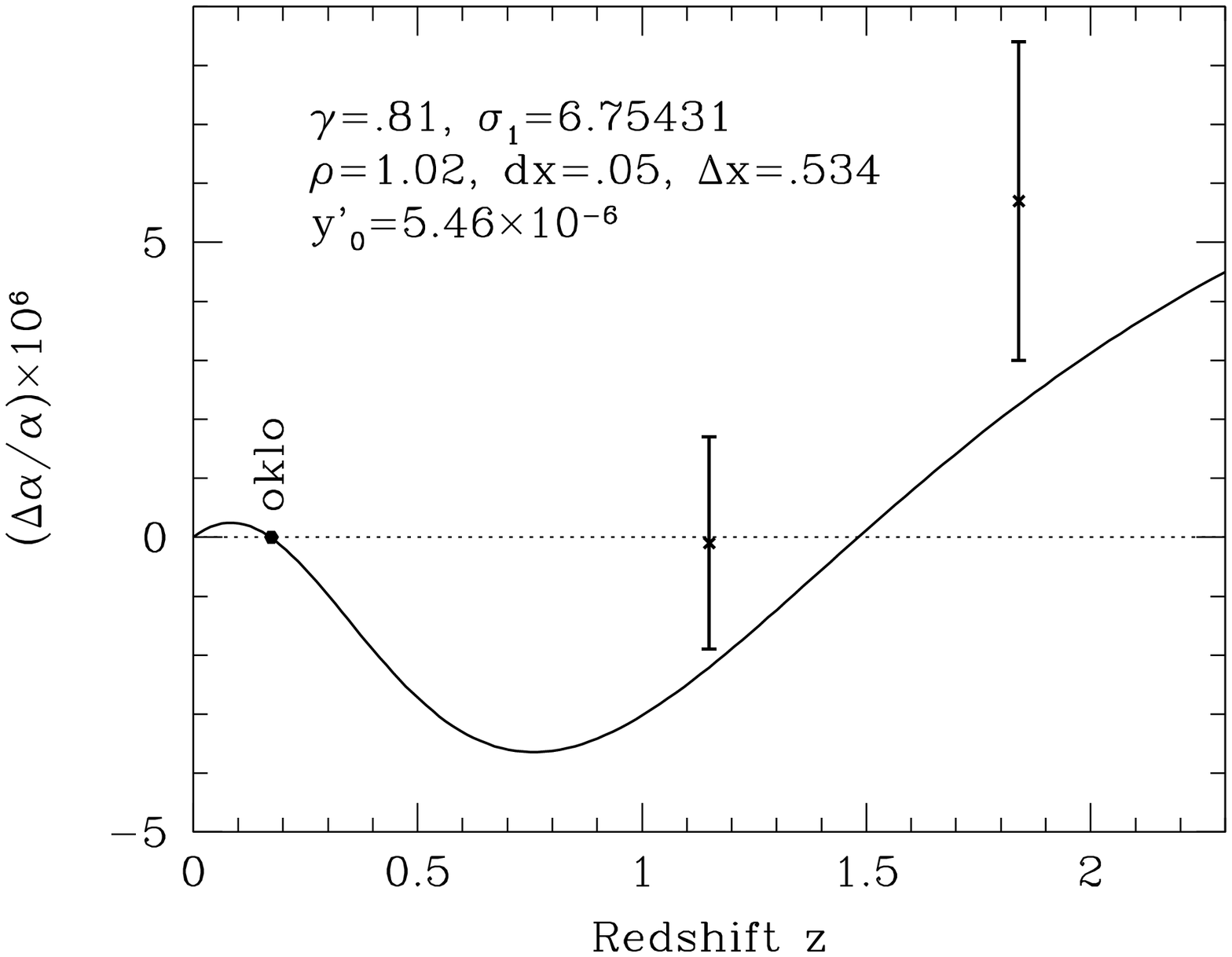}
\mbox{}\\[-.6em]
\eminip
\mbox{}\\[-2.4em]
\setcounter{figure}{8}
\begin{figure}[h]
\caption{An example of the ``desired" behaviors, instead of those in Fig. 1, obtained artificially by shifting the time coordinate $x$ in $\Delta\alpha/\alpha$ from the value responsible for the acceleration by $\Delta x=0.534$, agrees reasonably well with all of the observations, including  $v=0.4$.
}
\end{figure}
\mbox{}\\[-1.1em]

It might be still interesting to suggest that accepting curves as shown in 
Fig. 1 may not be the only way to fit the available data.  In fact Fig. 9 shows an example of the ``desired" behavior, which we constructed artificially by a mathematical trick to use $\sigma (x-\Delta x)$ in \reflef{tvalpha_1}) where $\Delta x=0.534$, by exploiting a low-frequency portion of the oscillating $\sigma$.  With $\sigma_1=6.75431$ we find that the curve quite different from those in Fig. 1 achieves reasonable agreements with all the observations including the laboratory experiment, though with the calculated $w$ even closer to $-1$ than in Fig. 6.  We expect to see how this behavior emerges in fact, by either changing the parameters and initial values in wider ranges, or modifying the potential slightly in (5.58) in \cite{cup}.

Searching for solutions beyond the reference solution might be suggested by yet another indication of the dark energy density to be present for $z \gsim 2$ for  the required growth of structure \cite{dage}.

\mbox{}\\
\noindent
{\large\bf Acknowledgments}\\[.6em]
I would like to thank Paolo Molaro for important comments on the QSO data, including the latest result before publication.  I also thank Shuntaro Mizuno and Matteo Viel for their discussions on the equation of state of dark energy.

\end{document}